\documentclass[aps,floatfix,prd,superscriptaddress,reprint,showpacs,10pt,preprintnumbers,longbibliography]{revtex4-1}
\usepackage[utf8]{inputenc}
\usepackage[pdftex]{graphicx}
\usepackage{subfigure}
\usepackage{float}
\usepackage{amssymb}
\usepackage{amsmath}  
\usepackage{dsfont}
\usepackage{color}
\usepackage{array}
\usepackage[locale=US]{siunitx}
\usepackage{bm}
\usepackage{mathrsfs}
\usepackage{pifont}
\usepackage{multirow}
\usepackage{tensor}
\usepackage{upgreek}
\usepackage[pdftex,
            pdftitle={O(3) with MPS},
            pdfauthor={Falk Bruckmann, Karl Jansen, Stefan Kuehn},
            bookmarks,
            colorlinks,
            linkcolor=myblue,
            citecolor=mymagenta,
            menucolor=black,
            urlcolor=myblue,
            plainpages=false,
            pdfpagelabels,
            hypertexnames=false]{hyperref}
\usepackage{verbatim}

\definecolor{mymagenta}{RGB}{200, 0, 100}
\definecolor{myblue}{RGB}{45, 48, 146}

\newcommand{\X}{\ding{53}}
\newcommand{\lmax}{\ensuremath{l_\mathrm{max}}}
\newcommand{\tr}{\ensuremath{\text{tr}}}
\newcommand{\ket}[1]{\mbox{$ | #1 \rangle $}}
\newcommand{\bra}[1]{\mbox{$ \langle #1 | $}}

\begin{document}
\title{O(3) nonlinear sigma model in 1+1 dimensions with matrix product states}
\author{Falk Bruckmann}
\affiliation{Universit{\"a}t Regensburg, Institut für Physik, Universit{\"a}tstraße 31, 93053 Regensburg, Germany}
\author{Karl Jansen}
\affiliation{NIC, DESY Zeuthen, Platanenallee 6, 15738 Zeuthen, Germany}
\author{Stefan K{\"u}hn}
\affiliation{Perimeter Institute for Theoretical Physics, 31 Caroline Street North, Waterloo, Ontario N2L 2Y5, Canada}

\date{\today}
\begin{abstract}
We numerically study the spectral properties, the entanglement and the zero-temperature phase structure at nonvanishing chemical potential of the O(3) nonlinear sigma model. Using matrix product states, a particular kind of one-dimensional tensor network state, we show that we are able to reach the asymptotic scaling regime and to reproduce the analytical predictions for the mass gap at vanishing chemical potential. In addition, we study the scaling of the entanglement entropy towards the continuum limit obtaining a central charge consistent with 2. Moreover, our approach does not suffer from the sign problem and we also explore the phase structure of the model for nonzero chemical potential and map out the location of the transitions between different charge sectors with high precision.
\end{abstract}

\preprint{DESY 18-213}

\maketitle
\section{Introduction}
Nonlinear sigma models in 1+1 dimensions with their spin/rotor degrees of freedom have a long history as interesting models in their own right and as benchmark models for four-dimensional gauge theories. Analogies include a negative beta function, which leads to the phenomenon of asymptotic freedom as well as nonperturbative effects like mass generation and topology. Moreover, the sign problem that hampers importance sampling at nonzero chemical potentials~\footnote{The interplay of chemical potentials and the related twisted boundary conditions leads to excitations of fractional topological charge~\cite{Eto:2004rz,Eto:2006mz,Bruckmann:2007zh,Brendel:2009mp,Harland:2009mf,Bruckmann:2018rra}, which are the basis for refined resurgence conjectures in sigma models~\cite{Dunne:2012zk,Dunne:2012ae,Cherman:2013yfa,Misumi:2014jua,Dunne:2015ywa}.} also occurs in sigma models. Dual variables have been shown to solve the sign problem in CP($N-1$) as well as O($N$) sigma models~\cite{Bruckmann:2015sua}, and subsequent simulations of the O(3) model~\cite{Bruckmann2015,
Bruckmann2016} revealed a second-order phase transition where the lowest mass in the spectrum equals the chemical potential, and with a dynamical critical exponent consistent with 2 as well as two-particle phase shifts in agreement with the analytical $S$-matrix~\cite{Zamolodchikov:1977nu}.

In this study we numerically examine the O(3) model in one spatial dimension in its Hamiltonian lattice formulation. One of our motivations is that in the Hamiltonian framework the introduction of a chemical potential does not cause additional difficulties in numerical simulations; it simply multiplies the Hermitian charge operator. Thus, we can directly address the zero-temperature physics, e.g., the ground state in each sector of fixed charge. We note in passing that in principle also nonzero temperature physics can be addressed in the Hamiltonian setup. A difficulty arises from the fact that the Hilbert spaces per site, which we represent in angular momentum eigenfunctions (spherical harmonics), are infinite dimensional. Hence, we have to truncate them to a finite dimension, which we do by restricting the total angular momentum at each site. Nevertheless, the total dimension grows exponentially with the volume, i.e., the number of sites, thus rendering numerical approaches dealing with the full Hilbert space infeasible. Here we use matrix product states (MPS) -- a particular kind of one-dimensional tensor network (TN) state -- which efficiently parametrize a subspace of weakly entangled states. In particular, MPS techniques allow for computing ground states and low-lying excitations for many physically relevant Hamiltonians, and the effort only grows polynomially in the size of the system and the tensors~\cite{Hastings2007,Verstraete2004}.

Most notably, numerical methods based on MPS do not suffer from the sign problem and directly yield the ground-state wave function at the end of the computation. This makes it possible to compute interesting (local) observables and to examine the entanglement structure in the state. The power of MPS methods for computing spectral properties~\cite{Byrnes2002,Sugihara2004,Banuls2013,Buyens2013,Rico2013,Kuehn2014,Shimizu2014,Silvi2014,Buyens2015,Buyens2016,Buyens2017,Milsted2016,Zapp2017,Banuls2017,Sala2018,Banuls2018,Banuls2018a,Sala2018a}, even in regimes which are inaccessible with Monte Carlo methods~\cite{Silvi2016,Banuls2016b,Banuls2016a,Banuls2016c}, thermal states~\cite{Saito2014,Banuls2015,Saito2015,Banuls2016} and simulating dynamical problems~\cite{Buyens2013,Kuehn2015,Pichler2015,Buyens2016b} for (1+1)-dimensional lattice field theories, has already been successfully demonstrated. However, most of the work so far has focused on gauge models which are not asymptotically free.

In this work we apply MPS to the asymptotically free O(3) nonlinear sigma model and  explore its ground state and the mass gap. Compared to previous TN studies of the model~\cite{Unmuth-Yockey2014,Kawauchi2015,Kawauchi2016,Kawauchi2016a,Unmuth-Yockey2017a}, we show that we are able to reach the asymptotic scaling regime, and, even for modest truncations, we do reproduce the analytical predictions, in particular in the scaling regime towards the continuum limit. This is rather nontrivial since in this regime large angular momenta are not suppressed, which will eventually render our truncations insufficient to faithfully describe the physics of the model when the gap closes exponentially.

Furthermore, we have access to the (bipartite) entanglement entropy in the ground state, and we investigate its scaling towards the continuum limit. As a consequence of the asymptotic freedom, the gap closes in this limit and the model becomes critical. We confirm the theoretically predicted logarithmic divergence with the correlation length~\cite{Calabrese2004} and extract the value for the central charge which we find to be consistent with 2.

Taking advantage of the fact the MPS methods do not suffer from the sign problem, we also explore the zero-temperature phase structure of the model at nonvanishing chemical potential. We observe the expected transitions between the different charge sectors of the Hamiltonian and are able to precisely locate the transition points.

The rest of the paper is organized as follows. In Sec.~\ref{sec:model} we introduce the Hamiltonian lattice formulation and the observables we are using, and describe our MPS approach. Subsequently, we present our numerical results in Sec.~\ref{sec:results}. Finally we discuss our findings in Sec.~\ref{sec:conclusion}. Technical details of our basis and numerical extrapolations are provided in Appendixes \ref{app:matrix_elements_truncation} and \ref{app:numerical_details}.

\section{Model and methods\label{sec:model}}
The simplest representation of the O(3) model is by its continuum Euclidean action,
\begin{align}
 S=\frac{1}{2g^2}\int\! d^2x\,
 \partial_\nu \mathbf{n}\,\partial_\nu \mathbf{n}
\label{eq:action}
\end{align}
where $\mathbf{n}$ is a real three-component unit vector, $g$ is the dimensionless (bare) coupling and $\nu$ is summed over one spatial dimension and Euclidean time extending from $0$ to the inverse temperature. As usual, the partition function is given by a path integral of $\exp(-S)$ over $\mathbf{n}$-field configurations.

On a lattice with spacing $a$, the two-dimensional integral over the derivative terms is replaced by $a^2$ times the $(x,\nu)$-sum over hopping terms $-2\mathbf{n}(x)\mathbf{n}(x+a\hat{\nu})/a^2$ (approaching $-\mathbf{n}\,\partial_\nu^2\mathbf{n}$ in the limit of vanishing lattice spacing), with the prefactor $\beta$ replacing $1/g^2$. Thanks to asymptotic freedom, the continuum limit is achieved by $\beta\to\infty$.

The action with chemical potential coupled to rotations in the first two components of $\mathbf{n}$ is obtained by replacing $\partial_\nu\to\partial_\nu-\delta_{\nu,0}\,\mu\,\tau_2$ with $\tau_2$ being the second Pauli matrix (see, e.g., Ref.~\cite{Bruckmann2014}) and, thus, is no longer real.

\subsection{Hamiltonian lattice formulation}
To study the model numerically with MPS, we switch to the Hamiltonian formulation. It can be deduced from the action above, but is also well known in the literature~\cite{Hamer:1978uq,Hamer1979,Shigemitsu:1980tx,Gliozzi1985,Hasenfratz1990,Hasenfratz:1990ab,Sachdev2007}. The potential term, corresponding to the spatial gradient, is discretized as described above. The kinetic term per site is that of a three-vector with fixed radius, i.e., momentum squared without radial component, which is nothing but the square of the orbital angular momentum. Hence, the lattice Hamiltonian for a system with $N$ sites reads 
\begin{align}
 aH &= \frac{1}{2\beta}\sum_{k=1}^N \mathbf{L}^2_k - a\mu\, Q -\beta \sum_{k=1}^{N-1} \mathbf{n}_k \mathbf{n}_{k+1},
 \label{eq:lattice_hamiltonian} \\
 Q&=\sum_{k=1}^N L^z_k.
\end{align}
In the expression above, $\mathbf{L}^2_k$ is the angular momentum operator acting on site $k$ (note the inverse $\beta$ in this kinetic term), and $\mu$ the chemical potential coupling to the total charge $Q$ which is nothing but the sum of third components of the angular momenta. Notice that contrary to conventional Monte Carlo approaches to the Lagrangian formulation we work with open boundary conditions for convenience in our MPS simulations~\footnote{Notice that MPS simulations are not restricted to open boundary conditions and there also exist algorithms for the periodic case~\cite{Verstraete2004}.}, and the sum for the potential term only ranges to $N-1$. Equation~\eqref{eq:lattice_hamiltonian} describes a linear chain of coupled quantum rotors, and operators acting on the same site fulfil the well-known commutation relations
\begin{align}
\begin{aligned}
 {[L^\alpha,L^\beta]} &= i\,\varepsilon^{\alpha\beta\gamma} L^\gamma,  & [L^\alpha,n^\beta] &= i\,\varepsilon^{\alpha\beta\gamma} n^\gamma, \\
 [n^\alpha,n^\beta] &= 0, & \alpha,\beta,\gamma&\in\{x,y,z\}.
\end{aligned}
\label{eq:commutation_relations}
\end{align}
The Hamiltonian conserves $Q$ because the kinetic term obviously commutes with $L^z$ at every site, and the commutator with the hopping term yields $[n_k^\alpha n_{k+1}^\alpha,\sum_{k'} L_{k'}^z]=-i\,\varepsilon^{\alpha z\gamma}(n_k^\alpha n_{k+1}^\gamma+(\alpha\rightleftharpoons \gamma))$, which vanishes as a result of the antisymmetry of the Levi-Civita symbol.

A suitable basis for such a system is the tensor product of the common eigenfunctions $|lm\rangle$ of $\mathbf{L}^2$ and $L^z$ for each site (whose spatial representations are the spherical harmonics),
\begin{align}
\begin{aligned}
 \mathbf{L}^2\ket{lm} &= l(l+1)\ket{lm}, &
 l&\in\mathds{N}_0^+,\\
 \quad\quad L^z\ket{lm} &= m \ket{lm}, & 
 m&\in[-l,l].
\end{aligned}
\label{eq:spherical_harmonics}
\end{align}
Note that the commutation relations from Eq.~\eqref{eq:commutation_relations} imply that the value of $l$ is not bounded from above and hence the basis is infinite dimensional. While the first two Hamiltonian terms are diagonal in this representation, the potential is more complicated. Defining the combinations $n^{\pm}=(n^x\pm i n^y)/\sqrt{2}$, we can rewrite the hopping part as $n_k^+ n_{k+1}^-+n_k^-n_{k+1}^++n_k^zn_{k+1}^z$. In Appendix~\ref{app:matrix_elements_truncation} we give the matrix elements $\bra{lm}n^{\pm}\ket{l'm'}$ and $\bra{lm}n^z\ket{l'm'}$ in terms of Wigner-3$j$ symbols. Selection rules are such that involved total angular momenta obey $|l'-l|=1$, while $m'$ agrees with $m\pm 1$ or $m$. In the Hamiltonian these expressions have to be used at neighboring sites $k$ and $k+1$. 

As we elaborate below, these states are very useful in the strong-coupling limit at small $\beta$, where the angular momentum term dominates. Near the continuum at large $\beta$, however, the hopping term tends to align the rotors. For this picture, independent angular momentum states are far from being effective, and it would be very useful to find a more suitable basis. 

\subsection{Observables and expectations\label{sec:observables_expectation}}
In the Hamiltonian framework, the charge $Q$ is simply the sum over all $m$ quantum numbers along the chain. The mass gap is essentially given by the energy difference between the first excited and the ground state, $am=a\Delta E\sqrt{\eta}=(aE_1-aE_0)\sqrt{\eta}$. Here we take into account the correction $\eta$ due to the anisotropy of spatial and temporal couplings in the Hamiltonian formalism~\cite{Shigemitsu1981}, to be able to make contact with the continuum prediction. From the one-loop renormalization one finds $\eta = 1/(1-1/\beta\pi)$. In the Lagrangian formulation, Eq.~\eqref{eq:action}, with periodic boundary conditions the continuum limit of the mass gap $m$ in units of the lattice spacing $a$ reads~\cite{Hasenfratz1990,Shigemitsu1981}
\begin{align}
 am  &=\frac{8}{e}\,a\Lambda_{\overline{\text{MS}}} = 64a\Lambda_L\notag\\
 &=128\uppi \beta\exp(-2\uppi\beta)\qquad (\beta\to\infty)
 \label{eq:gap}
\end{align}
where we have used the two-loop expression for the Callan–Symanzik beta function to compute $\Lambda_L$.

An advantage of the MPS approach to be used is that it allows for easy access to the entanglement entropy of the system. Dividing the system into two contiguous subsystems $A$ and $B$, we can efficiently compute the entanglement entropy for the bipartition which is given by the von Neumann entropy of the reduced density matrix
\begin{align}
 S= - \tr(\rho_A\log \rho_A),\quad\rho_A=\tr_B(\ket{\psi}\bra{\psi}).
\end{align}
In the expression above $\ket{\psi}$ is the state of the system and $\tr_B$ refers to the partial trace over the subsystem $B$. In our simulations, we analyze the entropy in the ground state and choose the subsystem to be half of the chain. 

For a fixed value of $\beta$ and vanishing $\mu$ the Hamiltonian from Eq.~\eqref{eq:lattice_hamiltonian} is local and gapped. Thus we expect the entanglement entropy $S$ for a subsystem $A$ to grow with its surface area~\cite{Hastings2007}. Since for a one-dimensional chain the latter consists of two points, $S_A$ is expected to saturate upon increasing the size of $A$.

Moreover, from Eq.~\eqref{eq:gap} we see that as we approach the continuum limit $\beta\to\infty$ the gap closes, and the system becomes critical. Hence, in the asymptotic scaling regime we expect the entropy for the reduced density matrix describing half of the system to diverge logarithmically as $S= (c/6)\log(\xi/a) +\bar{k}$, where $\xi/a$ is the correlation length in lattice units and $c$ the central charge of the underlying conformal field theory describing the critical point and $\bar{k}$ a (nonuniversal) constant~\cite{Calabrese2004}. Using the fact that the correlation length is inversely proportional to the gap, $\xi/a\propto 1/am$, we get together with Eq.~\eqref{eq:gap}
\begin{align}
S = \frac{c}{6}\left(2\pi\beta - \log\beta\right) + k,
\label{eq:entropy_vs_beta}
\end{align}
where $k$ sums up the constant terms. 

\subsection{Spectrum at strong coupling}
In the limit of small $\beta$ the system prefers zero angular momentum. Therefore, the ground state has  energy $E_0=0$. The first excited state (at $\mu=0$) has $l=1$ at a single site and thus $E_1=1/\beta+\mathcal{O}(\beta)$, which is also the leading order gap at small $\beta$ (the next to leading order correction is $-2\,\beta/3$~\cite{Hamer:1978uq,Hamer1979}).

If we enforce a nonzero charge, the angular momentum cannot vanish everywhere, since at least a single site has to have a nonzero quantum number $m$ and, thus, the associated $l$ must be nonzero, too. It is not hard to see that it is energetically favorable to induce a certain charge $Q$ (smaller than or equal to $N$) by assigning the minimal quantum numbers $(l,m)=(1,1)$ to $Q$ sites and $(l,m)=(0,0)$ to the complementary ones~\footnote{In other words, states with $(l,m)=(1,0),\,(1,-1)$ and $l=2,\,3$ etc.\ cost additional energy.}. The leading kinetic term is not sensitive to the spatial arrangement of those occupied sites, which yields ${N \choose Q}$ degenerate ``unperturbed eigenstates'', but the subleading hopping term knows about neighbors, and thus breaks this degeneracy: Let $L=\{0,1\}$ denote the relevant quantum numbers $(l,m)=\{(0,0),(1,1)\}$. Then the matrix elements of the hopping term, $\bra{L_k L_{k+1}}\mathbf{n}_k\mathbf{n}_{k+1}\ket{L'_k L'_{k+1}}$, are unity if $L_k=L'_k+1$ and $L_{k+1}=L'_{k+1}-1$, i.
e. if an excitation is created at site $k$ and annihilated at site $k+1$, or vice versa. This can be read as a \emph{Hubbard model}.

For periodic boundary conditions, it is possible to analyze the strong-coupling limit further analytically, at least for a small number of sites. In the fixed charge sector $Q=1$, for instance, degenerate perturbation theory on the $L\in\{0,1\}$ states reveals that the ground state is the sign coherent superposition $(\ket{10\ldots0}+ \ket{010\ldots0}+ \ldots+ \ket{0\ldots01})/\sqrt{N}$ with entanglement entropy $\log 2$. At $Q=2$ the ground-state coefficients increase with the distance between the two occupied sites. For a small number of sites $N=4,\,6,\,8,\,10$ we have determined the half-chain entropies to grow like $1.202,\,1.300,\,1.331,\,1.344$. For similar observations in the O(2) model at small $\beta$ see~\cite{Yang2016a,Unmuth-Yockey2017,Bazavov2017,Unmuth-Yockey2017a}.

\subsection{Phase structure for nonvanishing chemical potential}
The Hamiltonian conserves the total charge $Q$; hence it is block diagonal and each block can be labeled with the corresponding eigenvalue $q$ of the charge operator. For convenience, let us rewrite Eq. \eqref{eq:lattice_hamiltonian} as
\begin{align}
aH = -a\mu\, Q + aW_\mathrm{aux},
\end{align}
where $aW_\mathrm{aux}$ sums up the potential term and the kinetic part, and is independent of $\mu$. If we now restrict $aH$ to a block characterized by $q$, the charge operator is simply given by $q\mathds{1}$ and the Hamiltonian further simplifies to 
\begin{align}
aH|_q = -a\mu q \mathds{1} +a W_\mathrm{aux}|_q.
\end{align}
From this equation we can see that the ground-state energy inside a block with charge $q$ is given by
\begin{align}
aE_{0,q}(\mu) = -a\mu q + aE_0[aW_\mathrm{aux}|_q]
\label{eq:energy_vs_mu}
\end{align}
with $aE_0[aW_\mathrm{aux}|_q]$ being the minimum eigenvalue of the operator $aW_\mathrm{aux}$ restricted to block $q$ and independent of $a\mu$. Thus, inside a sector the ground-state energy scales linearly with the chemical potential and the slope is given by $q$. In particular, for zero charge the energy does not depend on $a\mu$ at all, which is also called Silver blaze property~\cite{Cohen2003}. Moreover, since $Q$ is proportional to the identity  inside a specific charge sector, the ground state for each block is independent of $a\mu$ and simply given by the ground state of $aW_\mathrm{aux}|_q$. 

While the energies $aE_{0,q}(\mu)$ can be measured in our approach, in the physics of grand-canonical ensembles only the chemical potential is fixed, and the system minimizes these energies over all charge sectors. The global minimum of the energy selects a certain charge sector that depends on $a\mu$. At certain values $a\mu_c$ the energy levels of two sectors with different charges cross and it is energetically favorable to go from one sector with charge $q$ to another one with charge $\bar{q}$, which generically is the successor, $\bar{q}=q+1$. 
The value of $a\mu_c$ can be determined by equating $aE_{0,q}(\mu_c)=aE_{0,q+1}(\mu_c)$, which yields
\begin{align}
a\mu_c = aE_0[aW_{\text{aux}}|_{q+1}]-aE_0[aW_{\text{aux}}|_{q}].
\label{eq:transition_location}
\end{align}
Thus, we expect the total charge of the ground state to exhibit discontinuous changes as we increase the chemical potential. 

In particular, the first transition, from 0 to unit charge, is expected to happen at the energy gap, $a\mu_c=a\Delta E$ (related to the mass $am$ as discussed in Sec.~\ref{sec:observables_expectation}). This follows from the equation above, if in the absence of $\mu$, i.e.\ for $aW_\mathrm{aux}$, the ground state has $Q=0$, $aE_0[aW_\mathrm{aux}]=aE_0[aW_\mathrm{aux}|_0]$, whereas in the first excited state the massive particles form a triplet and $Q=1$, being the lowest state in that sector, $aE_1[aW_\mathrm{aux}]=aE_0[aW_\mathrm{aux}|_1]$. As a result, the right-hand side of Eq.~\eqref{eq:transition_location} is nothing but the energy gap. The next transition can be used to extract two particle energies in finite volumes and from them phase shifts~\cite{Bruckmann2015}.

In the infinite-volume limit all these transitions merge to one curve for the charge \emph{density} as a function of $\mu$, and only the first critical $\mu_c=m$ survives to mark -- at zero temperature -- a second order quantum phase transition~\cite{Bruckmann2016}. Similar observations have been made in the O(2) model at small temperature~\cite{Banerjee2010}.

\subsection{Truncation in angular momentum}
In our Hamiltonian, Eq.~\eqref{eq:lattice_hamiltonian}, the spectrum of $\mathbf{L}^2$ (and $L^z$) is unbounded; thus the Hilbert space even for a single lattice site is infinite dimensional. In any numerical approach working on a finite basis one therefore has to apply a truncation. We choose to keep all basis states at all sites with 
\begin{align}
 l_k \leq \lmax \quad (\forall\text{ sites } k)\,.
 \label{eq:l_truncation}
\end{align}
Together with all possible $m$ quantum numbers this yields a local Hilbert space with dimension $d=\sum_{l=0}^{\lmax}(2l+1)=(\lmax+1)^2$.

Note that the normalization of $\mathbf{n}$ and the commutator relations from Eq.~\eqref{eq:commutation_relations} are violated when representing $\mathbf{n}$ in such a truncated basis. As we show in Appendix~\ref{app:matrix_elements_truncation}, this only concerns expectation values where the highest sector with $\lmax$ is involved. Nevertheless, the truncated Hamiltonian still conserves the total charge $Q$. Thus, for large enough values of $\lmax$ the violations are expected to be negligible, which will be checked \emph{a posteriori}.

\subsection{Numerical approach}
In order to compute the low-lying spectrum of the Hamiltonian from Eq. \eqref{eq:lattice_hamiltonian} we use the MPS ansatz. For a system with $N$ sites on a lattice with open boundary conditions the ansatz reads
\begin{align}
\ket{\Psi} = \sum_{i_1,\dots,i_N=1}^d M_1^{i_1} M_2^{i_2} \dots M_N^{i_N} \ket{i_1}\otimes \dots \otimes \ket{i_N}
\end{align}
where $\ket{i_k}$, $i_k=1,\dots ,d$ is a basis for the $d$-dimensional Hilbert space on site $k$. The $M_k^{i_k}$ are $D\times D$-dimensional complex matrices for $1<k<N$ and $M_1^{i_1}$ ($M_N^{i_N}$) is a $D$-dimensional complex row (column) vector. The parameter $D$, called the bond dimension of the MPS, determines the number of variational parameters in the ansatz and limits the amount of entanglement that can be present in the state (see Refs.~\cite{Verstraete2008,Schollwoeck2011,Orus2014a,Bridgeman2017} for detailed reviews). For our system, $i_k$ should be read as a super index for $(l_k,m_k)$ with the summation ranges discussed above.

In our simulations we are interested in the ground state and the energy gap of the model. The MPS approximation for the ground state can be found variationally by minimizing the energy. To this end, one iteratively updates the tensors $M_k^{i_k}$, one at a time, while keeping the others fixed~\cite{Verstraete2004}. The optimal tensor in each step is found by computing the smallest eigenvalue~\cite{Stathopoulos2010,Wu2016} of the effective Hamiltonian describing the interactions of site $k$ with its environment. After obtaining the ground state, the first excited state can be found in a similar fashion by projecting the Hamiltonian on a subspace orthogonal to the ground state and running the same algorithm with the projected Hamiltonian~\cite{Banuls2013,Banuls2013a}.

\section{Results\label{sec:results}}
\subsection{Spectral properties}
Let us first focus on the case of vanishing chemical potential. In order to benchmark our approach, we study the ground-state energy and the energy gap for a range of values of $\beta\in[0.65;1.8]$ and different system sizes $N=40,80,120$. In addition, to probe for the effects of truncating the local bases to a finite angular momentum, we explore $\lmax=1,2,3,4$. Moreover, we have another source of error due to the limited bond dimension available in the numerical simulations. This error can be controlled by repeating the calculation for every combination of $(\beta,\lmax,N)$ for several bond dimensions $D\in[80,160]$ and extrapolating to the limit $D\to\infty$ (see Appendix~\ref{app:numerical_details} for details on the extrapolation procedure).

\begin{figure}[htp!]
\centering
\includegraphics[width=0.48\textwidth]{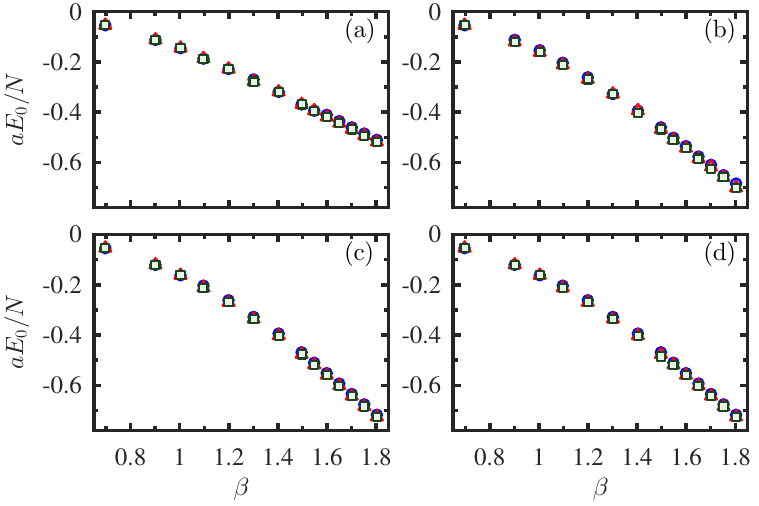}
\caption{Ground-state energy density as a function of $\beta$ for $\lmax=1$ (a),  $\lmax=2$ (b), $\lmax=3$ (c),  and $\lmax=4$ (d) and system sizes $N=40$ (blue dots), $N=80$ (red triangles) and $N=120$ (green squares). The error bars from the extrapolation in $D$ are smaller than the markers.}
\label{fig:GS_energy}
\end{figure}

In Fig.~\ref{fig:GS_energy} we show the extrapolated results for the ground-state energy density $aE_0/N$ for various values of $\lmax$ and $N$. In general, we observe that the errors due to the finite bond dimension in our simulation are negligible and there is almost no dependence on the system size, as the values for $aE_0/N$ for $N=40,80$ and $120$ are essentially identical. For the simplest nontrivial truncation, $\lmax=1$, we see that the ground-state energy density notably differs from those obtained for larger values. In contrast, there is hardly any difference between the values obtained with $\lmax\geq 2$, only for larger values of $\beta$ the results for $\lmax=2$ deviate slightly from those for $\lmax=3,4$. 

\begin{figure}[htp!]
\centering
\includegraphics[width=0.48\textwidth]{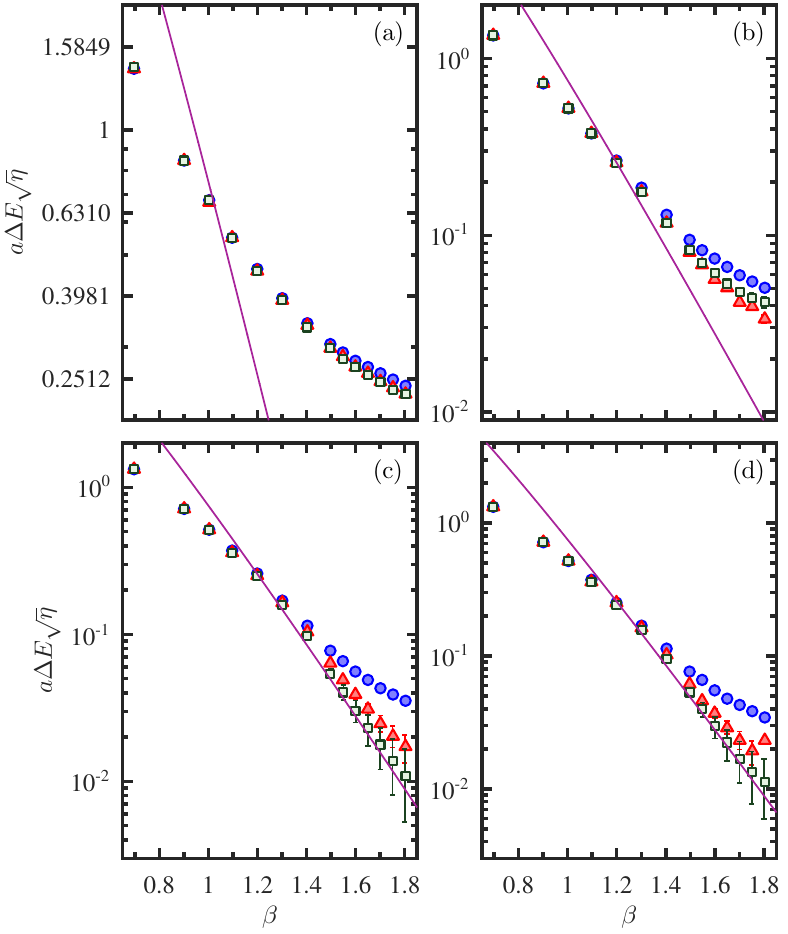}
\caption{Mass gap as a function of inverse coupling for $\lmax=1$ (a),  $\lmax=2$ (b), $\lmax=3$ (c),  and $\lmax=4$ (d) and system sizes $N=40$ (blue dots), $N=80$ (red triangles) and $N=120$ (green squares). The solid purple line shows the theoretical prediction for the asymptotic scaling of the gap according to Eq.~\eqref{eq:gap}.}
\label{fig:gap}
\end{figure}

While the ground-state energy density is fairly insensitive to the truncation and does not show strong finite-size effects, the situation is noticeably different for the energy gap. Figure~\ref{fig:gap} reveals that there is a significant difference between results for different truncations with $\lmax\leq 3$, and only for the largest two values of $\lmax$ our results are in agreement. Although finite-size effects are small for $\lmax=1,2$, we see from Figs.~\ref{fig:gap}(a) and \ref{fig:gap}(b) that our numerical values for the gap in those cases are not compatible with the asymptotic scaling predicted by Eq.~\eqref{eq:gap}. These data rather seem to approach a constant value as the slope decreases with increasing values for $\beta$.  On the contrary, for $\lmax=3,4$ there are much stronger finite-size effects. While for all our system sizes the data seem to enter the asymptotic scaling around $\beta\approx 1.2$~\footnote{At this value of $\beta$ the factor $\sqrt{\eta}$ is $1.17$ and decreases further towards $1$ for increasing $\beta$.}, the ones for $N=40,80$ eventually start to deviate from the theoretical prediction around $\beta=1.4$. Only for our largest system size, $N=120$, we recover the asymptotic scaling up to the largest value of $\beta=1.8$ we study, as can be seen in Figs.~\ref{fig:gap}(c) and \ref{fig:gap}(d). In addition, we observe that in this case our error bars are growing as the value of $\beta$ increases. This is a direct consequence of the asymptotic scaling: since the gap closes exponentially with increasing $\beta$ and we use the same range of $D$ values in all our simulations, the MPS approximations we obtain are becoming progressively worse. Nonetheless, the errors are reasonably small up to $\beta=1.6$ and we reliably recover the asymptotic scaling between $1.2\leq \beta \leq 1.6$.

\subsection{Scaling of the entanglement entropy towards the continuum limit}
\begin{figure}[htp!]
\centering
\includegraphics[width=0.48\textwidth]{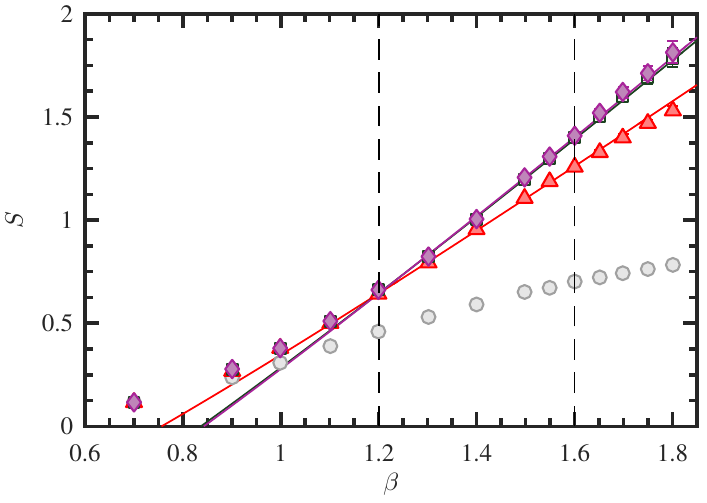}%
\caption{Half-chain entropy as a function of the inverse coupling for $N=120$ and $\lmax=2$ (red triangles),  $\lmax=3$ (green squares), and $\lmax=4$ (magenta diamonds). The solid lines indicate a fit to our data according to Eq.~\eqref{eq:entropy_vs_beta} inside the window indicated by the vertical dashed lines. For completeness, we also show the entropy for the case of  $\lmax=1$ (gray circles).}
\label{fig:entropy}
\end{figure}

Taking advantage of the fact the we have easy access to the entanglement entropy  in the ground state, we can also explore its scaling towards the continuum limit. Our results for the gap show that only for our largest system size the finite-size effects are small enough that we recover the asymptotic scaling regime over such a large range of values for $\beta$. Hence, for the following we focus on $N=120$ and study the half-chain entropy and its scaling as we approach the continuum limit. Our results are shown in Fig.~\ref{fig:entropy}. 
For $\lmax=3,4$ we clearly observe an almost linear scaling for large $\beta$, as predicted in leading order by Eq.~\eqref{eq:entropy_vs_beta}, and there is hardly any difference between the data for both truncations. The entropies for $\lmax=2$  still show an approximately linear behavior towards the continuum limit; however, the values for larger $\beta$ are significantly smaller than those obtained for $\lmax=3,4$. For completeness, we also show the data for the simplest nontrivial truncation, $\lmax=1$. As one can see, these do not show the linearly divergent behavior in $\beta$ but rather seem to approach a constant value, thus giving an indication that the model does not become critical as we approach the continuum. In particular, this is also in agreement with our observation in Fig.~\ref{fig:gap}(a) that the gap is not closing.

Fitting our data for $\lmax\geq 2$ to the theoretical prediction, we can extract the central charges, too. Looking again at our results for the gap in Figs.~\ref{fig:gap}(c) and \ref{fig:gap}(d), we see that we enter the asymptotic scaling regime around $\beta=1.2$ and that our errors are reasonably small up to $\beta=1.6$. Hence, we choose this range to fit our data for the entropy to Eq.~\eqref{eq:entropy_vs_beta}. Additionally, we estimate our systematic error by comparing the results from different fitting intervals (see Appendix~\ref{app:numerical_details} for the details). The results for the central charge obtained by this procedure 
are shown in Table~\ref{tab:central_charge}.
\begin{table}[htp!]
\begin{tabular}{cc}
\hline
$\lmax$ & $c$  \\ \hline \hline
2 & $1.66 \pm 0.19$ \\  
3 & $2.01 \pm 0.12$ \\
4 & $2.04 \pm 0.14$ \\ \hline
\end{tabular}
\caption{Central charges extracted from the scaling of the entanglement entropy. The error represents a systematic uncertainty related to the choice of our fitting interval.}
\label{tab:central_charge}
\end{table}
For $\lmax=2$, we see that our data points for $\beta>1.6$ are progressively below our fit, thus indicating that the slope further decreases as we go closer to the continuum limit. Hence, our value of $c=1.66$ for that case only seems to be an upper bound. Similar to $\lmax=1$ this might give an indication that the model does not become critical, consistent with our observation in Fig.~\ref{fig:gap}(b) that the gap does not follow Eq.~\eqref{eq:gap} but rather seems to tend to a constant value. On the contrary, for $\lmax=3,4$ our fit describes the data in the entire range for $\beta$ we study and we obtain values of $c\approx 2$.  Note that O($N$) models in the perturbative regime, for which the nonlinearity does not play a role, describe $N-1$ massless bosons. These turn into $N$ massive bosons at low energies for $N\geq 3$. Thus, the measured central charge of the ground state points to the two perturbative degrees of freedom.

These observations together with the gap provide a comprehensive picture of the effect of the truncation. Although the ground-state energy density is rather insensitive to the truncation in angular momentum, we observe that for $\lmax=1,2$ the gap does not seem to close, and the model does not become critical for $\beta\to\infty$, in particular for $\lmax=1$. Only for $\lmax=3,4$ we capture the relevant features and recover the theoretical prediction for the asymptotic scaling of the gap as well as the divergence in the entropy according to Eq.~\eqref{eq:entropy_vs_beta} as we approach the continuum limit. 

\subsection{Phase structure in the presence of a chemical potential}
Contrary to conventional Monte Carlo methods, our MPS approach does not suffer from the sign problem and we can also investigate the phase structure of the model at nonvanishing chemical potential. To this end we explore the ground-state energy and the total charge as a function of $a\mu$ for $\beta=1.0, 1.1, 1.2$ and $N=40, 60, 80$, which is around the beginning of the asymptotic scaling for $a\mu=0$, but still far away from regimes with noticeable finite-size effects. Again, we study several truncations $\lmax=1,2,3,4$ and bond dimensions $D\in[80;240]$ to estimate the errors of the truncation as well as our numerical errors.

\begin{figure}[htp!]
\centering
\includegraphics[width=0.48\textwidth]{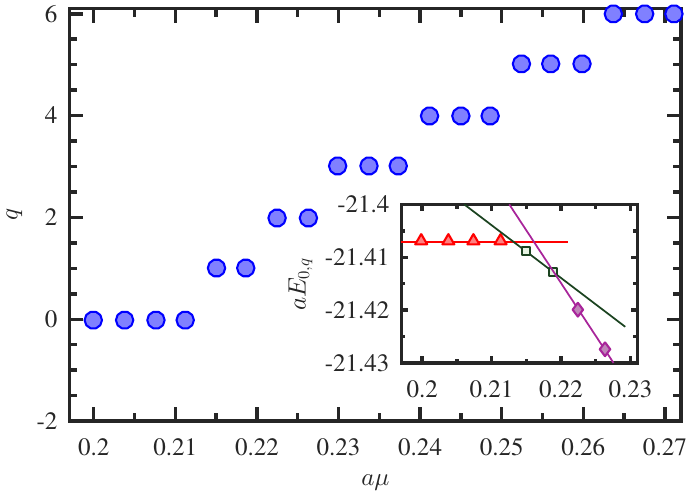}%
\caption{Charge as a function of the chemical potential for $N=80$, $\beta=1.2$, $\lmax=4$ and $D=200$. Inset: Corresponding ground-state energies for data points in the sector for $q=0$ (red triangles), $q=1$ (green squares) and $q=2$ (purple diamonds). The solid lines represent the theoretical prediction according to Eq.~\eqref{eq:energy_vs_mu}, where the slope is fixed by the charge value and $aE_0[aW_\mathrm{aux}|_q]$ has been determined from the numerical data.}
\label{fig:q_vs_mu}
\end{figure}
Figure~\ref{fig:q_vs_mu} shows an example of the total charge in the ground state as a function of the chemical potential. Indeed, we see the theoretically predicted discontinuous changes in the values of $q$ by one unit as $a\mu$ is growing. In addition, the inset in Fig.~\ref{fig:q_vs_mu} shows that the behavior of the ground-state energy is in excellent agreement with the theoretical prediction from Eq.~\eqref{eq:energy_vs_mu}, and we observe a linear scaling with a slope given by the total charge of the sector.

\begin{figure*}[htp!]
\centering
\includegraphics[width=0.95\textwidth]{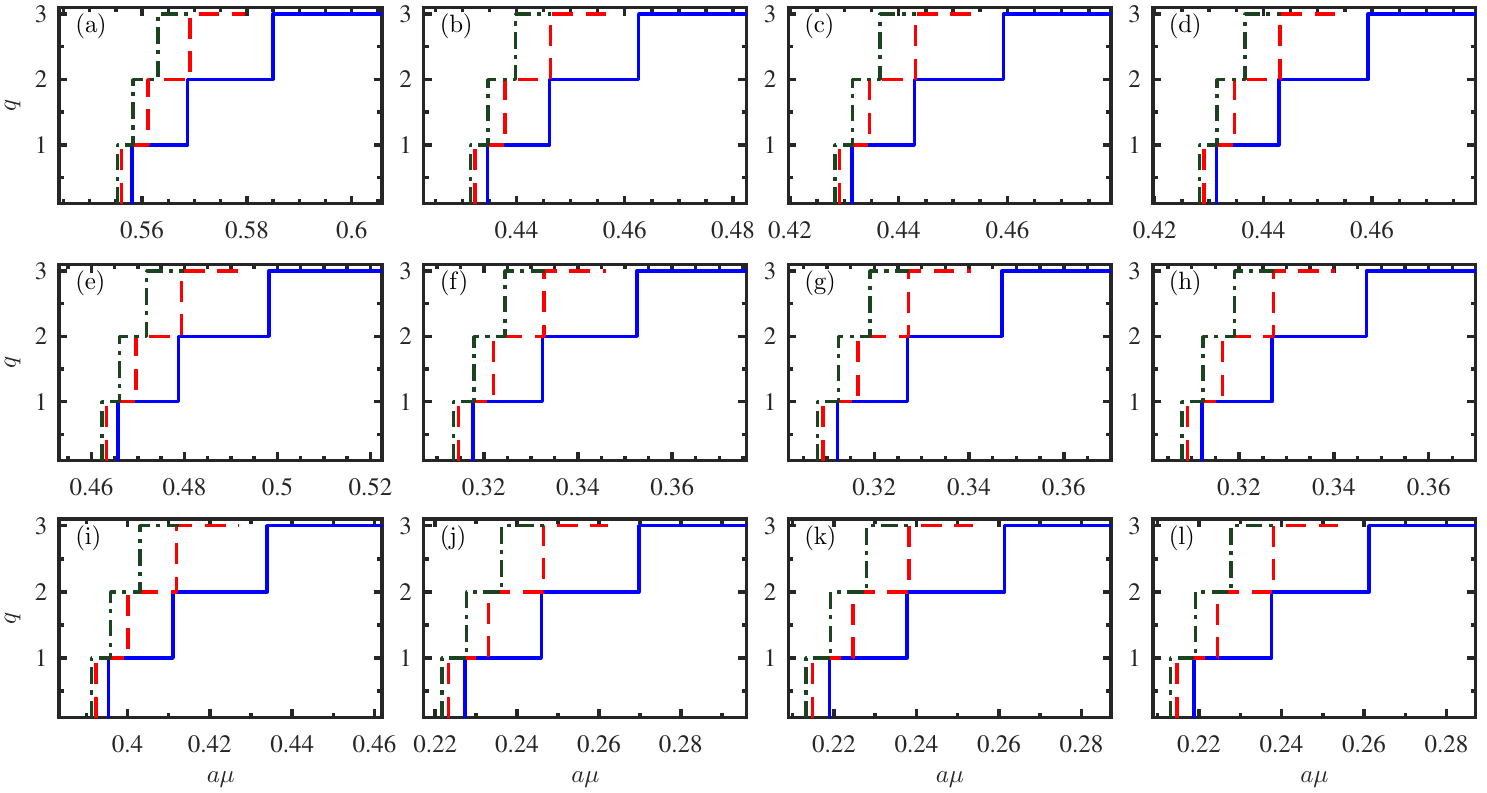}%
\caption{Charge as a function of the chemical potential for $\beta=1.0$ (first row), $\beta=1.1$ (second row), $\beta=1.2$ (third row) and $\lmax=1$ (first column), $\lmax=2$ (second column), $\lmax=3$ (third column), $\lmax=4$ (fourth column). The different lines correspond to different system sizes, $N=40$ (blue solid line), $N=60$ (red dashed line), $N=80$ (green dash-dotted line).}
\label{fig:phase_structure}
\end{figure*}

Similar to the case of vanishing chemical potential, we can extrapolate our results for the ground-state energies to the limit $D\to\infty$, and, using Eqs.~\eqref{eq:energy_vs_mu} and \eqref{eq:transition_location}, we can estimate the exact location of the transitions (see Appendix~\ref{app:numerical_details} for details). Our results for the various truncations, system sizes and couplings are shown in Fig.~\ref{fig:phase_structure}.
In general, we observe that our data do not show very strong truncation effects and there is hardly any difference between results with $\lmax\geq 2$. Only for the simplest nontrivial truncation, $\lmax=1$, the locations of the transitions are significantly shifted towards higher values of the chemical potential. Comparing data with different values of the coupling, we see that the transitions between different charge sectors occur for smaller values of $a\mu_c$ as we go closer to the continuum, independently of the truncation and the volume. Moreover, for a fixed value of $\beta$ the widths of the plateaus in each charge sector shrink with increasing system size, consistent with the expectation that in the infinite-volume limit the plateaus for $q>0$ merge, and there is only a single second-order quantum phase transition. 

In particular, independent of the truncation the first transition should occur at $a\mu_c = a\Delta E$. In Table~\ref{tab:transition} we compare the values for both quantities obtained in our simulations for $N=80$. 
\begin{table}[htp!]
\centering
\begin{tabular}{|c||c|c|}
\hline
\multicolumn{3}{|c|}{$\lmax=1$} \\ 
$\beta$ & $a\mu_c$ & $a\Delta E$ \\ \hline
$1.0$ & $0.55528 \pm	5.00222\times 10^{-12}$ & $0.55509 \pm	0.00012$\\ 
$1.1$ & $0.46230 \pm	5.13980\times 10^{-11}$ & $0.46219 \pm	0.00007$\\ 
$1.2$ & $0.39124 \pm	3.60401\times 10^{-10}$ & $0.39119 \pm	0.00003$\\ 
\hline \hline
\multicolumn{3}{|c|}{$\lmax=2$} \\ 
$\beta$ & $a\mu_c$ & $a\Delta E$ \\ \hline
$1.0$ & $0.43145 \pm	5.89100\times 10^{-09}$ & $0.43113 \pm	0.00021$\\ 
$1.1$ & $0.31350 \pm	5.14500\times 10^{-08}$ & $0.31337 \pm	0.00008$\\ 
$1.2$ & $0.22163 \pm	6.73692\times 10^{-07}$ & $0.22158 \pm	0.00003$\\ 
\hline \hline
\multicolumn{3}{|c|}{$\lmax=3$} \\ 
$\beta$ & $a\mu_c$ & $a\Delta E$ \\ \hline
$1.0$ & $0.42823 \pm	9.06657\times 10^{-08}$ & $0.42800 \pm	0.00017$\\ 
$1.1$ & $0.30800 \pm	3.59573\times 10^{-06}$ & $0.30789 \pm	0.00007$\\ 
$1.2$ & $0.21320 \pm  2.84339\times 10^{-06}$ & $0.21313 \pm	0.00004$\\ 
\hline \hline
\multicolumn{3}{|c|}{$\lmax=4$} \\ 
$\beta$ & $a\mu_c$ & $a\Delta E$ \\ \hline
$1.0$ & $0.42823 \pm	9.06657\times 10^{-08}$ & $0.42800 \pm	0.00016$\\ 
$1.1$ & $0.30794 \pm	7.69692\times 10^{-06}$ & $0.30783 \pm	0.00007$\\ 
$1.2$ & $0.21307 \pm	3.15277\times 10^{-06}$ & $0.21299 \pm	0.00004$\\ 
\hline
\end{tabular}
\caption{Chemical potential $a\mu_c$ for the first transition for $N=80$ and various truncations. For comparison we show our values for the mass gap obtained for that case at $a\mu=0$. The errors represent the systematic uncertainty due to our extrapolation procedure (see Appendix~\ref{app:numerical_details} for details).}
\label{tab:transition}
\end{table}
Indeed, the critical $a\mu_c$ at the first transition is in almost perfect agreement with the energy gap. We recapitulate that the latter has been obtained at the same parameters $(\beta,\lmax,N)$ but with $a\mu=0$ by an independent computation, determining the lowest energy in the Hilbert space orthogonal to the ground state.

\section{Conclusion\label{sec:conclusion}}
We have studied the spectral properties, the entanglement entropy in the ground state and the zero-temperature phase structure at nonvanishing chemical potential for the O(3) nonlinear sigma model using MPS. To render the Hilbert space finite dimensional, we have restricted the angular momentum at each site to a certain maximum value $\lmax$. 

Looking at the ground-state energy density at vanishing chemical potential, we observe that it is rather insensitive to the truncation, and does not show strong finite-size effects. We find that $\lmax\geq 2$ is already sufficient to avoid noticeable effects even in the weak-coupling regime. In contrast, the energy gap is much more sensitive to the truncation. For $\lmax=1,2$ the gap does not show the expected asymptotic scaling, but rather seems to approach a constant value as we go towards the continuum limit, independently of the volume. Increasing $\lmax$ to $3$, $4$ we observe that for large enough volumes the gap closes exponentially in the weak-coupling regime, as predicted by perturbation theory.

Investigating the scaling of the bipartite entanglement entropy in the ground state towards the continuum limit, we observe a similar picture as for the mass gap. For $\lmax=1$ we do not recover the expected divergence as we approach the continuum limit and the model does not become critical, consistent with the observation that the mass gap does not close according to the theoretical prediction. For $\lmax=2$ our data do not allow us to fully rule out that the energy gap closes and the model becomes critical. However, the central charge obtained in this case is significantly smaller than those for larger $\lmax$ and the deviations become more pronounced the closer we go to the continuum limit. On the contrary, for $\lmax=3,4$ our data for the entropy show the predicted divergence, and the central charges obtained in those cases are consistent with $2$.

Our MPS approach does not suffer from the sign problem and readily allows us to investigate the zero-temperature phase structure at nonvanishing chemical potential. We clearly observe the theoretically predicted transitions between different charge sectors of the Hamiltonian, and we can determine the location of the transitions with great precision. Here truncation effects are only clearly visible for $\lmax=1$, and the transition points are significantly shifted towards larger values of the chemical potential compared to results with $\lmax\geq 2$, which show hardly any difference among each other. With increasing volume the transition points between phases with charges $q\geq 1$ come closer to each other, consistent with the expectation that in the infinite-volume limit only single second-order quantum phase transition survives. In particular, we have verified that the first transition occurs at chemical potential values equal to the gap, independently of the truncation and the coupling.

Our findings suggest that even moderate values of $\lmax$ are sufficient to capture the relevant features of the model, at least for the ground state and the phase structure at nonvanishing chemical potential. Similar observations have been made in MPS studies for O(2) and O(4) rotor models~\cite{Milsted2016} as well as for simple (1+1)-dimensional gauge models~\cite{Buyens2017,Banuls2017}. This is especially encouraging for potential future quantum simulators for rotor models~\cite{Zou:2014rha}. Although current quantum simulation experiments for lattice field theories~\cite{Martinez2016,Muschik2016,Kokail2018,Hartung2018} are rather limited, one might nevertheless be able to study relevant phenomena even with a modest amount of resources. Moreover, our results can serve as a test bench for these experiments. 

Concerning the phase structure of the model, more features of the ground state at high chemical potentials can be investigated. As the system tends to become planar~\cite{Bruckmann2016} it leans towards the O(2) model, but the conjectured Berezinskii-Kosterlitz-Thouless transition of the latter~\cite{Bruckmann2014} has not been seen in numerics~\cite{Bruckmann2016}. Secondly, the induced particles form a massive triplet (with known $S$-matrix~\cite{Zamolodchikov:1977nu}), which could modify entanglement properties and consequently the central charge. 

On the technical side, it would also be very intriguing to find another basis for the weak-coupling regime such that truncation effects become much milder. This would allow a much improved investigation of the continuum limit. Finally, it would be interesting to probe the potential effects of spatially twisted boundary conditions which are crucial for resurgence conjectures of sigma models~\cite{Dunne:2012zk,Dunne:2012ae,Cherman:2013yfa,Misumi:2014jua,Dunne:2015ywa}.

\begin{acknowledgments}
We thank Ashley Milsted, Christine Muschik and Norbert Schuch for helpful discussions. This research was supported in part by Perimeter Institute for Theoretical Physics. Research at Perimeter Institute is supported by the Government of Canada through the Department of Innovation, Science and Economic Development Canada and by the Province of Ontario through the Ministry of Research, Innovation and Science. F. B. acknowledges support by DFG (Contract No.\ BR 2872/9-1).

Computations were made on the supercomputer Mammouth Parallèle 2  from the University of Sherbrooke, managed by Calcul Québec and Compute Canada. The operation of this supercomputer is funded by the Canada Foundation for Innovation (CFI), the ministère de l'Économie, de la science et de l'innovation du Québec (MESI) and the Fonds de recherche du Québec - Nature et technologies (FRQ-NT). 
\end{acknowledgments}

\appendix

\section{Basis formulation and truncation to a finite dimension\label{app:matrix_elements_truncation}}
Here we show how to calculate the matrix elements for the components of the operator $\mathbf{n}$ in the angular momentum eigenbasis. The calculation can be simplified by realizing that the operators $n^\pm$, $n^z$ in position representation are related to the spherical harmonics with $l=1$, which in turn form a spherical tensor operator of rank 1, as follows:
\begin{align}
\begin{aligned}
 \begin{pmatrix}
 X_{-1} \\ X_0 \\ X_1
 \end{pmatrix}
 &:=
 \begin{pmatrix}
 (x-iy)/\sqrt{2} \\ z \\ (-x-iy)/\sqrt{2}
 \end{pmatrix}\\
 &=\sqrt{\frac{4\pi}{3}}
 \begin{pmatrix}
 Y_{1,-1} \\ Y_{1,0} \\ Y_{1,1}
 \end{pmatrix}
 \quad\quad (r=1)\,.
\end{aligned}
\end{align}
Thus the matrix elements can be computed with the Wigner–Eckart theorem and the expectation values become integrals over three spherical harmonics,
\begin{align}
 \bra{lm}X_M\ket{l'm'}
 =(-1)^m\sqrt{\frac{4\pi}{3}}\!\int \!\!d\Omega\: Y_{l,-m}\,
 Y_{1,M}\,Y_{l',m'}
\end{align}
where we have used that $Y_{l,m}^* = (-1)^mY_{l,-m}$. The integral is related to Wigner-3$j$ symbols (which are proportional to Clebsch-Gordan coefficients) by a well-known formula and we obtain
\begin{align}
\begin{aligned}
 \bra{lm}X_M\ket{l'm'}
 &=(-1)^m\sqrt{(2l+1)(2l'+1)}\\
 &\times
 \begin{pmatrix}l & 1 & l' \\0 & 0 & 0\end{pmatrix}
 \begin{pmatrix}l & 1 & l' \\-m & M & m'\end{pmatrix}
\end{aligned}
\end{align}
The right-hand side is nonzero if and only if the conservation $m=m'+M$ holds and 
$|l'-l|=1$ (as a consequence of the triangle relation, however, $l'\neq l$, from the first symbol). The expectation values needed in this work are obtained by identifying $n^{\pm}\rightleftharpoons\mp X_{\pm 1}$ and $n^z \rightleftharpoons X_0$.

As discussed in the main text, Eq.~\eqref{eq:l_truncation}, we truncate the local Hilbert space through limiting $l$ by $\lmax$ at every site. Consequently, the normalization of $\mathbf{n}$ and the commutator relations from Eq.~\eqref{eq:commutation_relations} will be violated, as we demonstrate now. The truncation means that we represent all operators by finite dimensional matrices,
\begin{align}
 L^\alpha
 &\to \sum_{l_1,l_2}^{\lmax}\:\sum_{m_1, m_2}
 \ket{l_1m_1}
 \underbrace{\bra{l_1m_1}L^\alpha\ket{l_2m_2}}
 _{\textstyle \sim\delta_{l_1l_2}} 
 \bra{l_2m_2}\\
 n^\alpha
 &\to \sum_{l_1,l_2}^{\lmax}\:\sum_{m_1, m_2}
 \ket{l_1m_1}
 \bra{l_1m_1}n^\alpha\ket{l_2m_2}
 \bra{l_2m_2}
\end{align}
where the $n^\alpha$-expectation values have been calculated above. 
For bilinears of $n^\alpha$ we have to take the matrix product; e.g., the normalization becomes
\begin{align}
\begin{aligned}
 n^\alpha n^\alpha
 \to &\sum_{l_1,l_2}^{\lmax}\:\sum_{m_1, m_2}
 \ket{l_1m_1}\bra{l_2m_2}\\
 &\times\sum_{l'}^{\lmax}\sum_{m'}\bra{l_1m_1}n^\alpha\ket{l'm'}
 \bra{l'm'}n^\alpha\ket{l_2m_2}
\end{aligned}
\label{eq:app_operator_product}
\end{align}
Obviously, if $l'$ were summed up to infinity, the second line upon completeness would simply yield $\delta_{l_1,l_2}\delta_{m_1,m_2}$. From above we know that $l'$s up to $\max(l_1+1,l_2+1)$ contribute to such sums of expectation values, but they are neglected in our truncation once $l_{1}$ or $l_2$ equal $\lmax$. Thus, the normalization is violated in the highest sectors of angular momentum. This violation does not affect lower $l$-sectors or operators made of just $L$'s, since their expectation values are nonzero only within the same $l$-sector; so larger $l'$ are irrelevant anyhow.

That the truncated Hamiltonian still conserves the total charge $Q$ follows from a similar argument. Looking at the derivation below Eq.~\eqref{eq:commutation_relations}, we have to check the commutator of $n^\alpha$ with $L^z$ at some site (the neighboring site being a spectator) in the truncated representation. In the corresponding expectation values, as in the second line of Eq.~\eqref{eq:app_operator_product}, the $L^z$ factor again limits the relevant $l'$ to $l_{1,2}$, which are kept such that the completeness is intact and $[L^\alpha,n^\beta] = i\,\varepsilon^{\alpha\beta\gamma} n^\gamma$ holds indeed.

\section{Numerical details\label{app:numerical_details}}
Here we discuss some details of the evaluation procedures for the numerical data and how we estimate our errors for the data shown in the main text.

\subsection{Ground-state energy, mass gaps and ground-state entropy}
\begin{figure}[htp!]
\centering
\includegraphics[width=0.48\textwidth]{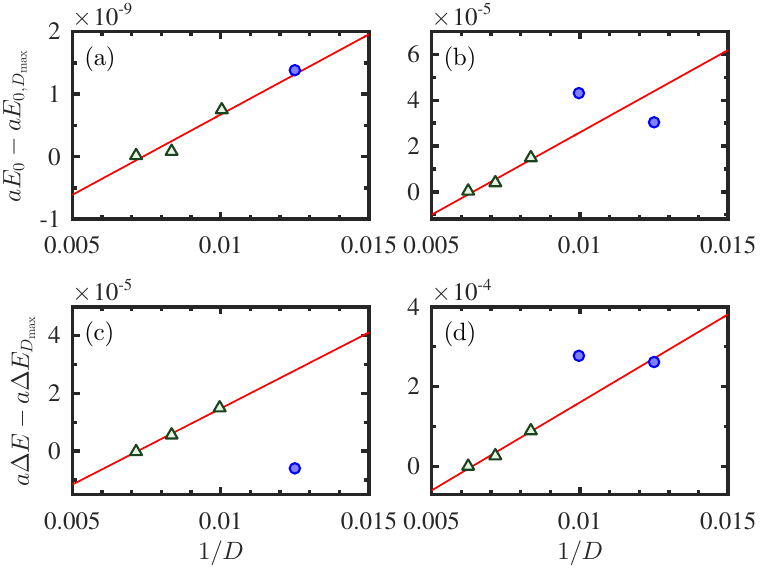}%
\caption{Extrapolation in bond dimension for the ground-state energy (upper row) and the mass gap (lower row) for $N=80$, $\beta=1.3$, $\lmax=1$ (left column) and $\lmax=4$ (right column). The green triangles indicate the data points used for the linear extrapolation (solid line).}
\label{fig:extrapolation_nomu}
\end{figure}

As we already mentioned in the main text, for every combination of system  size $N$, coupling $\beta$ and truncation $\lmax$ we repeat the numerical simulation for a range of bond dimensions $D\in[80,160]$. Subsequently we can estimate the exact value and the numerical error for the ground-state energy and the mass gap by extrapolating to limit $D\to\infty$. To this end, we extrapolate our data points for $aE_0$ and $a\Delta E$ with the largest three bond dimensions linearly in $1/D$ (see Fig.~\ref{fig:extrapolation_nomu} for an example). As an estimate for the central value we take the mean value of our data point with the largest bond dimension and the extrapolated value. The error is estimated as half of the difference between those two values.

\begin{figure}[htp!]
\centering
\includegraphics[width=0.48\textwidth]{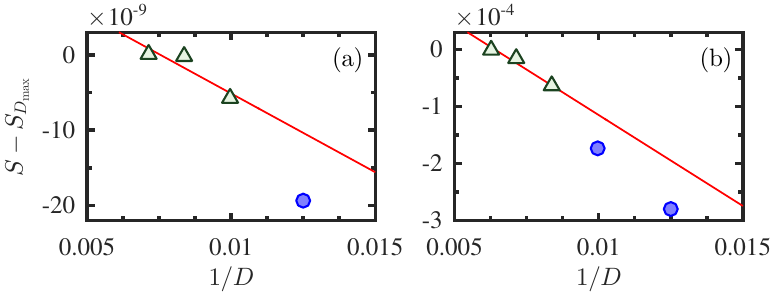}%
\caption{Half-chain entropy in the ground state as a function of bond dimension for $N=80$, $\beta=1.3$, $\lmax=1$ (a) and $\lmax=4$ (b).}
\label{fig:entropy_gs_vs_D}
\end{figure}

For the half-chain entropies of the ground state we proceed in the same fashion (see Fig.~\ref{fig:entropy_gs_vs_D} for an example). Similar to the ground-state energy, we observe that our results are well converged in bond dimension and the numerical errors due to limited values of $D$ in our simulations are negligible.

The errors for the central charges in Table~\ref{tab:central_charge} are estimated by comparing the results of different fits. As explained in the main text, the central value is obtained by fitting our data to Eq.~\eqref{eq:entropy_vs_beta} in the interval $1.2\leq \beta \leq 1.6$. Subsequently we repeat the same fit in every possible interval $[\beta_\mathrm{min},\beta_\mathrm{max}]$ with $1.2\leq\beta_\mathrm{min} <\beta_\mathrm{max}\leq 1.8$, which contains at least six data points. As an estimate for the error we take the maximum absolute value of the difference between the results from these fits and our central value. 

\subsection{Extracting the location of the transitions}
The location of the transitions can be determined with the help of Eq.~\eqref{eq:transition_location} from the main text. In our simulations we measure the ground-state energy and the charge. Hence, we can extract $aE_0[aW_\mathrm{aux}|_q]$ using Eq.~\eqref{eq:energy_vs_mu}. Combining these two equations, we find for the location of two adjacent phases with $q+1$ and $q$
\begin{align}
a\mu_c = aE_{0,q+1}(\bar{\mu})+a\bar{\mu}(q+1) - \left(aE_{0,q}(\tilde{\mu}) +a\tilde{\mu}q\right)
\label{eq:transition_location2}
\end{align}
where $\bar{\mu}$ and $\tilde{\mu}$ refer to the values of the chemical potential at which we determined the constants $aE_0[aW_\text{aux}|_{q+1}]$ and $aE_0[aW_\text{aux}|_{q}]$ for the two phases. Notice that Eq.~\eqref{eq:transition_location2} allows us to extract the location of the transition between two adjacent phases as long as we have a single data point in each of them available. Thus, this method is more efficient than extracting the transition points directly from the discontinuities of $Q$ as we do not need a high resolution in $\mu$ to determine them precisely.

\begin{figure}[htp!]
\centering
\includegraphics[width=0.48\textwidth]{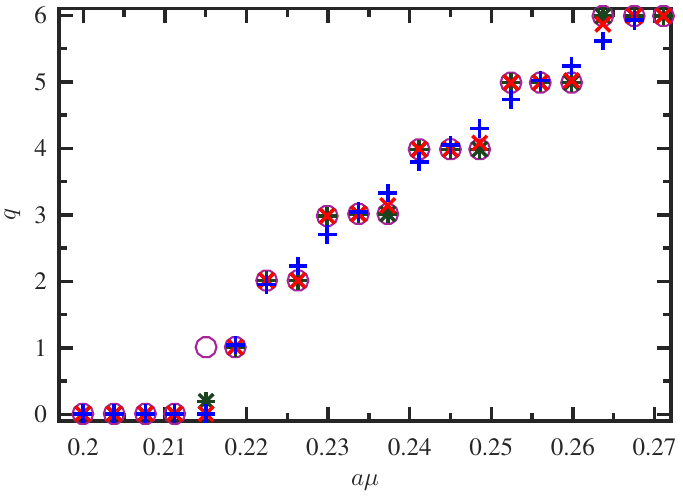}%
\caption{Charge as a function of the chemical potential for $N=80$, $\beta=1.2$, $\lmax=4$ and bond dimensions $D=80$ (blue crosses), $120$ (red \X's), $160$ (green asterisks), and $200$ (purple circles).}
\label{fig:q_vs_mu_several_D}
\end{figure}

\begin{figure}[htp!]
\centering
\includegraphics[width=0.48\textwidth]{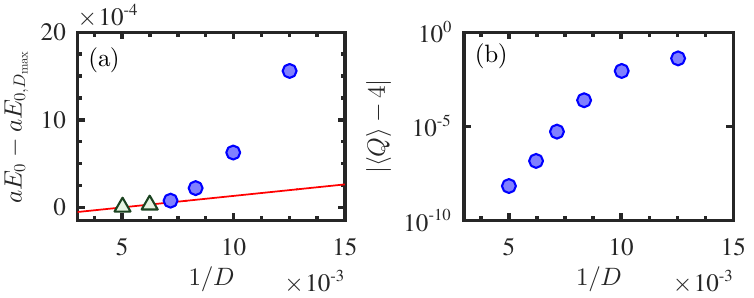}%
\caption{Ground-state energy as a function bond dimension (a) and deviation of the total charge from the exact integer value (b) for $N=80$, $\beta=1.2$, $\lmax=4$ $a\mu=0.2449$, which is in the sector with $q=4$ (see also Fig.~\ref{fig:q_vs_mu_several_D}). The green triangles in the left panel indicate the values used for the linear extrapolation which is shown by the red line.}
\label{fig:ext_D_q_dev}
\end{figure}

The value of $q$ can in principle be determined exactly, since the total charge is still an exact symmetry of the truncated Hamiltonian. In practice, finite bond dimension effects can break this symmetry and it is only restored for large enough values of $D$, as the example in Fig.~\ref{fig:q_vs_mu_several_D} shows.
Hence, in our simulations we start with $D=80$ and keep increasing the bond dimension until the deviations of $q$ from integer values are negligible. For all the cases we study we find that a bond dimension of $240$ is sufficient to stabilize the values of $q$ and to allow us to unambiguously determine in which phase we are [see Fig.~\ref{fig:ext_D_q_dev}(b) for an example].

The exact values for the ground-state energy and the corresponding errors are again estimated analogous to the case of vanishing chemical potential by extrapolating linearly in the last two data points [see Fig.~\ref{fig:ext_D_q_dev}(a) for an example].
From our extrapolation we obtain an estimate for the exact ground-state energy as well as the systematic error $\delta aE_{0,q}$. Propagating this systematic error in Eq.~\eqref{eq:transition_location2}, we obtain the systematic error for the transition points, which is given by
\begin{align}
\delta a\mu_c =\left|\delta aE_{0,q}(\bar{\mu})\right|+ \left|\delta aE_{0,q}(\tilde{\mu})\right|.
\end{align}
In general, we observe that this systematic error due to the finite bond dimension in the simulations is rather negligible and for all data shown in the main text the error bars are smaller than the markers (see also Table~\ref{tab:transition} in the main text).

\bibliography{Papers}
\end{document}